\documentclass[11pt,twoside]{article}

\usepackage{asp2006}
\usepackage{epsf}
\usepackage{lscape}

\begin{document}
\title{Creation and use of Citations in the ADS}  
\author{Alberto Accomazzi, G\"unther Eichhorn, Michael J. Kurtz, 
Carolyn S. Grant, Edwin Henneken, Markus Demleitner, Donna Thompson, 
Elizabeth Bohlen, Stephen S. Murray}
\affil{Harvard-Smithsonian Center for Astrophysics, 60 Garden St., Cambridge, MA, 02138, USA}

\keywords{astronomical data bases: miscellaneous}

\begin{abstract} 

With over 20 million records, the ADS citation database is regularly 
used by researchers and librarians to measure the scientific impact of 
individuals, groups, and institutions.   In addition to the traditional 
sources of citations, the ADS has recently added references extracted from 
the arXiv e-prints on a nightly basis.  We review the procedures used 
to harvest and identify the reference data used in the creation of 
citations, the policies and procedures that we follow to avoid 
double-counting and to eliminate contributions which may not be 
scholarly in nature.  Finally, we describe how users and institutions 
can easily obtain quantitative citation data from the ADS, both 
interactively and via web-based programming tools.

The ADS is available at http://ads.harvard.edu.

\end{abstract}

\section{Introduction}

With the introduction of citation data in 1997, the 
Smithsonian/NASA Astrophysics Data System (ADS) abstract service
started offering its users the option of finding out what
papers are citing a particular work, and, similarly, what 
are the works referenced by it.  As predicted, the addition
of these data has proven to be very popular among our users,
and as our coverage of citations has expanded from the core
astronomy journals into the field of physics and e-prints,
the use of the ADS citations has steadily increased.

In addition to the traditional source of citations provided by the
Science Citation Index by the Institute for Scientific Information
(ISI), the last few years have seen the advent of new sources of
bibliographic services offering citation information, among them
Google Scholar and Citebase.  With so many different options
available, it is natural for librarians and scientists
to wonder if the ADS should be considered the ultimate source 
for bibliometric data in astronomy or if other sources should be
consulted as well.  
This paper tries to give a nuanced answer to the question, discussing 
how the ADS maintains its citations and in which ways it differs
from other bibliographic databases.

In the first part of this 
paper we describe in detail the sources of citation
data which the ADS relies upon and the procedures we have in place
to help us maintain our citation lists up-to-date and accurate,
discussing the issues related to completeness and coverage.
We then illustrate the many ways in which citations play an 
integral part of the ADS abstract service through a sampling
of different bibliometric applications readily available to
ADS users.
We conclude by offering some comments about the usefulness
of citations in the ADS and the effect they have on our system.

\section{Generating Citations in the ADS}

Occasionally we are
asked by our users why a certain citation appears to be missing
from the ADS citations list.  The answer to this question is best given by 
explaining how the ADS creates the list of citations to a particular paper.
The procedure is the following:

\begin{enumerate}
\item Scan the full-text of paper A and find the reference section
\item Identify the individual reference strings and parse bibliographic 
tokens in them
\item Create a tentative bibcode B for each reference string based on the parsed tokens
\item Verify the existence of bibcode B in the ADS and compute similarity score 
between the ADS record and parsed reference
\item If the score is high enough, then we say that the reference has been resolved 
to the record corresponding to B and we save the pair (A, B) in the resolved reference 
table (``A cites B"), otherwise we go back to step 2
\item Create citations by inverting the reference table (``B is cited by A")
\end{enumerate}

The process described above has been implemented as a set of 
automated procedures that on 
a daily basis scan any new reference data entered into the ADS database
and try to parse and resolve the data as best as they can.  The procedures 
used to perform these tasks are described in \citet{1999ASPC..172..291A}
and \citet{2004ccdm.conf..521D}.

Based upon this approach, it follows that a particular 
citation from paper A to paper B will be properly attributed if {\em all} 
of the following are true:

\begin{enumerate}
\item Citing paper A is in the ADS bibliographic database
\item The ADS has fulltext or reference section of citing paper A available for
analysis
\item The reference to paper B has been successfully parsed and identified
\item The cited paper B is in the ADS bibliographic database
\end{enumerate}

Each bibliographic record for which the ADS has obtained a reference list will
have a link to its list of references which have been identified as ADS records.
Conversely, papers which have been identified as having been referenced
in the literature will have a link to the papers citing them.

It is worth noticing that the list of references displayed by the ADS 
does not include any references which we were not able
to successfully match to an existing ADS record.  This approach 
differs from the one followed by ISI and Citebase, which instead attempt to
display the list of references as extracted from the fulltext with links to 
the individual records that have been successfully identified from the reference
strings.  While this approach may be regarded as more user friendly,
the ADS decided not to follow it.
One reason for this is that our agreements with the publishers 
providing us with the fulltext paper information often do not 
allow us to display the references published therein,
but rather want us to point the user to the fulltext article
available on their website.
Another issue of concern is the usefulness of
providing reference information that is not necessarily properly formatted or 
interpreted, as is often the case for the entries that cannot be resolved.
This difference in handling the display of references to the user does not 
however affect the attributions of citations since in all cases citations can
only be counted when the cited paper is successfully identified as an 
existing record.

\section{Sources of Citation Data}

The first source of
citation data became available to the ADS in 1997 when the AAS purchased 
citation data for the core astronomy journals from ISI.  These
consisted of about 1.3 million citations which 
constituted the original citation data provided by the ADS.
The list of citations from ISI was expanded and updated in 2000 and then 
again in 2003 to include more astronomy journals as well as some of the 
major physics journals, and it now includes more than 9 million
citations for the period 1982-2002.
In late 1999 we added over 1.2 million 
citations generated from processing digitized fulltext scans in our article 
archives using Optical Character Recognition (OCR) technology,
an effort which is still ongoing.  During that
same year we also received the first set of electronic records from the 
University of Chicago Press which included reference data and which were
integrated in the citation database.  The addition of more electronic
reference data from the other main publishers of astronomy journals in 2000 and
from the APS journals in 2001 brought the total number of citations above 
10 million.  During the past five years more publishers have been collaborating
with the ADS in the exchange of citation data, including IoP, Springer and 
EDP Sciences, bringing us to the current total of over 20 million resolved 
citations, generated by parsing 30 million references.  A summary of the major
contributors of reference data is available in Table 1.

\begin{table}[!ht]
\caption{Sources of Citation data in the ADS as of September 2006
and their coverage.
Please note that there is some overlap in these numbers.
The number of records listed (second column) corresponds to the references
successfully identified as ADS records.
The third column gives the success rate of the
reference resolution process for the particular dataset.
}
\smallskip
\begin{center}
\begin{tabular}{lrrl}
\tableline
\noalign{\smallskip}
Source & Records & Resolved & Date Range \\
       & (millions)  &  & \\
\noalign{\smallskip}
\tableline
\noalign{\smallskip}
APS          &  10.26 & 61\% & 1893-current \\
arXiv        &  7.50  & 61\% & 1992-current \\
Blackwell    &  0.92  & 81\% & 1999-current \\
EDP Sciences &  0.81  & 86\% & 2001-current \\
IoP          &  4.59  & 52\% & 1887-current \\
ISI          &  9.83  & 72\% & 1982-2002    \\
OCR          &  3.13  & 82\% & 1950-2002    \\ 
Springer     &  0.43  & 58\% & 1997-current \\
UCP          &  1.93  & 87\% & 1995-current \\
\noalign{\smallskip}
\tableline
\end{tabular}
\end{center}
\end{table}

As of March 2005, references from the arXiv e-prints
have been integrated in 
the ADS.  When we retrieve the metadata for the nightly update of arXiv 
e-prints, we also process the fulltext articles to extract the references 
contained in them.  As Table 1 shows, this source of references
has become an integral and important part of the ADS, supplementing the
data provided by the publishers.  
However, it also presents us with new challenges, the biggest one being a
lack of uniformity in the structure of the papers.  Since each e-print is
prepared for submission to one of several hundred different publications,
and since there is no copy-editing involved, both contents and formatting
of these papers may be hard to predict.  As a result, there is always 
the possibility that our procedures may be unable to successfully extract
references from the e-prints.

In addition to the technical challenges involved in obtaining references
from e-prints, a natural concern about their inclusion 
is whether these references will be duplicated once 
the paper is published and the ADS receives the corresponding record from
the publisher.  In order to avoid the duplication of citations,
we have been performing extensive
matching between the e-prints and published articles. 
All e-prints for which we find a matching journal article with 
associated references will not contribute to the citation count
of the cited papers.
Whenever a (journal) article is published for which we have the 
preprint in our system, we will replace the preprint references 
with the references from the paper, if they are available. 
If they are not, we continue to use the references from the 
preprint but attribute them to the published paper.

Another concern that has been raised about the inclusion of references
from e-prints has to do with the scholarly quality of works
posted to the arXiv.
As a precautionary measure against counting citations generated from 
e-prints that never make it into the published literature, we have
instituted the policy of excluding the references from a preprint
older than one year from the citation lists, unless the preprint has
been published.  Having adopted these safeguards, we feel that 
the integration of citations from e-prints and indeed the availability
of e-print records in the ADS has had a major, positive effect on research,
allowing our users to stay in touch with results from the latest
research more efficiently, without compromising their use of and access
to the refereed literature \citep{2005IPM....41.1395K}.

\section{Citation Completeness}

Due to the process involved in the creation of citations, and its
inherently imperfect nature,
we cannot guarantee even a certain degree of citation completeness,
and we do our best to assure that users are aware of this problem.
As discussed in the previous two sections, the
sources of citation incompleteness in the ADS may be summarized as follows:

\begin{itemize}

\item The ADS does not have the cited article in the database. 
This happens for instance for most papers appearing in mathematics, 
chemistry, and geophysics journals.
If you are somebody who publishes in any of these fields, we simply
will not know about those papers, much less about their citations.

\item Our reference resolver program could not interpret the reference. 
This may be due to errors or incompleteness in the reference, unusual 
formatting of the reference or the paper itself, 
or simply limitations in our program's abilities.
Given the many citation styles used in the literature, and the great
variety of formatting used by publishers supplying us with reference data,
this is not
a trivial matter to address, and often requires a fair amount of
training and supervision.

\item We do not have the reference list for the citing paper. 
This happens for older articles and for articles in journals and 
conference proceedings that do not supply us with reference lists.
Although this often comes as a surprise to some of our users, the fact
that we have a record for a particular paper does not mean that
we also know what are the articles that it has referenced.

\end{itemize}

We are constantly adding to our database by extracting reference lists 
from scanned articles, new electronic articles, and 
trying to improve reference recognition capabilities, so this
ongoing effort will cause the number of references and therefore citations
to increase over time.
We are also going back in time and adding back-records of
older papers published in astronomy and physics, so that more of the 
reference data in our system will be successfully matched to these papers
once they are entered in ADS.  We periodically check our reference lists to
see what publications are being cited in the core astronomy and physics 
journals and
which we are not able to successfully match against existing ADS records.
Based on this analysis, we can then quantitatively assess what are the most
referenced scholarly sources which do not appear in ADS yet, and work towards
creating new metadata for them.  As an example, Albert Einstein's famous 1905
papers were recently added to the ADS physics database after noticing the
many unresolved citations they were receiving in the published literature.

Despite our best efforts, we are well aware that we cannot always achieve 
the degree of citation completeness and accuracy that some people may desire, 
particularly in light of the importance that citations play
in today's world of academia.  As a result, we have been providing ways for
our users to submit corrections to citations already in our database
and supplement our own citation data with author- and librarian-provided 
bibliographic data.  The result of this effort 
is that we are now able to review and 
integrate user-submitted citation information in a timely manner.  
More information on 
how to submit this data to the ADS is available from our web site.

\section{Bibliometric Applications}

With the inclusion of citations in our system, researchers and librarians
have started using this data for a variety of bibliometric studies.  
Despite our warnings about the incompleteness of citation data in ADS
(and indeed the wisdom of using citation data in the first place),
the temptation to quantify the scientific output of a user or project
has proved too great for many people to resist.  As a result, we have 
provided services that allow our users to easily gain this information.
In the following sections, we briefly illustrate some of the most
popular and most requested ADS searches involving the use of citations,
offering some advice and warnings on their usefulness.
In all cases the first step is accessing one of the ADS abstract search 
interfaces, available at http://adsabs.harvard.edu/ads\_abstracts.HTML.
A word of caution: the procedures outlined below apply to the
capabilities available in the ADS as of September 2006.  
Being a system that is actively enhanced on a regular basis, readers 
should expect that some of the options described here may have
changed by the time this paper is read.

\subsection{Author Citations}

Using the standard ADS abstract search query, one can easily find out
the list of papers published by a particular author or group, or 
on a particular topic.  With the introduction
of citations, one can now further process these results so that the
most cited ones are listed first.  In this section we will show 
a few ways currently available in ADS for obtaining bibliometric 
information about a particular author or project.
In the subsequent examples, we have assumed that a person is looking
for citations to his or her own papers, but the procedures that we
illustrate can be similarly applied to a list of people or to one
of the bibliographic groups known to ADS and included in its abstract
service pages.

{\bf How many citations do I have?}
Query the ADS for all publications by you, making sure that the
number of returned records is adjusted so that they are all included
in the resulting list.  Then use the menu available on top of the resulting
page to resort the results by citations, which will display the most cited
papers at the top and will also give the total number of citations for the
set of papers.

{\bf How many of my citations are refereed?}
Using the list obtained at the previous step, go to the bottom of the form,
click on ``Select all Records,'' and then click on ``Get refereed citation
lists for selected articles.''  The resulting page contains all the refereed
papers that cite your papers, and the total number of the citations to
your papers is given at the top of the page.

{\bf How many citations have been made to my papers during the year X?}
Perform a query for your papers, select all records, 
and enter the year in question as both the
starting and ending year for the publication year range displayed in the 
menu at the bottom of the form.  Send off the query by clicking on
``Get citation lists for selected articles.''  
The total number of citations for the year in question is displayed
at the top of the page.

{\bf How do I exclude self-citations?}
After performing the original author query, select all records, 
and then click on ``Exclude self-citations'' at the bottom of the
result page and on ``Get citations lists for selected articles.'' 

{\bf What is my {\em h-index}?}
Recall that a scientist has an h-index (or Hirsch number) of N
if he or she has published N papers with at least N citations each.
The number can be easily obtained by
performing the first citation search illustrated above, and then going down the
list of papers noting the citation count and rank for each of them
until finding the one for which the rank is smaller or equal to the
citation count.  The rank of the paper in question is your N number.
More information on applications of the h-index metric 
using the ADS can be found in \citet{uta}.

\subsection{Keyword-based Searches}

Just as one can apply citation metrics to author queries, 
we can also use similar techniques on keyword-based searches.
In addition to obtaining citation counts, we
can also perform some additional follow-up queries of interest,
as outlined below (a more complete discussion of these queries 
can be found in \citealt*{2002SPIE.4847..238K}).
Let us assume that we are interested in a topic of interest
X (say ``the virtual observatory''); we start by querying
ADS for papers on the subject, and we obtain a list of papers
that are relevant to the subject.

{\bf What are the most cited papers on topic X?}
If we resort the list of papers returned by ADS on topic X by
citation counts we can immediately see what the most cited papers
are. However, one should be aware that citations accumulate 
during the lifetime of a paper, so that there is an age effect
in the ranking.  All things being equal, older papers will tend 
to have a greater number of citations than more recent ones
(simply because they have had more time to accumulate citations),
so one should keep this in mind when making comparisons.

{\bf What are the most useful papers on topic X?}
If we select all papers on our topic and then click on
``Get reference lists for selected papers,'' we obtain
the list of papers most cited by the original list of
records on our topic.  These can be considered most
useful to somebody who is interested in subject X,
because a large number of papers on the subject cite them,
and must therefore contain important information about it.
Note that the papers on this list are not necessarily on topic
X themselves.  Usually these are articles describing
surveys, instruments, techniques and theories that were
used by the scientists researching topic X.

{\bf What are the most instructive papers on topic X?}
Having obtained a list of interesting, recent or useful
papers on our topic of interest, we can now find the list
of papers which cite a large number of these articles, and
which therefore contain extensive coverage of topic X.
These can be considered the most instructive papers on the 
topic.  Often these are review articles on the subject,
and represent a good starting point for somebody new to the field.

By combining these types of searches with selections on date ranges 
and/or bibliographic groups, one can easily narrow the scope of
the search, for instance to consider only papers published in 
the last few years.

\subsection{Keeping up with Updates}

As discussed in the previous sections, scientists
and librarians will notice their number of citations change
over time, usually increasing as more publications reference
their works.  Similarly, as new e-prints and journal papers
appear in the literature, the rankings of papers on a particular 
topic will change.
In response to scientists' desire to stay current with research
developments, the ADS has recently introduced the {\em myADS}
personal notification service 
\citep{2003AAS...203.2005K,2006cs........8027H}.  
This service provides up-to-date 
customizable alerts to its subscribers whenever new bibliographic
records of interest to them are added to the ADS databases.
In particular, it allows a scientist to keep track of his/her
citations, show recent papers published by a list of authors
that he or she follows regularly, and provides a list of the
most recent, most popular and most cited papers on his or her 
topics of interest.
This service can be also useful to librarians or data providers
who maintain a list of bibliographies related to one or more projects,
since {\em myADS} can be configured through a series to queries
to return any new papers related to the project.
We strongly encourage active researchers and librarians to 
use this service in order to stay current with the latest 
developments in their research field.

\subsection{Automated Access}

Most of the queries that we have outlined above can be automated
by making use of simple web clients available in the public domain.  
Programmatic access to these queries is facilitated by the fact that ADS
offers the capability of generating output records in a few 
different highly-structured formats.  Among them are three different
XML formats, one of which (``XML Abstracts'') includes a citation
count for each paper returned by a query as well as a total citation count
for the entire set of papers (for a citation query).
Records for individual papers can similarly be downloaded 
as formatted XML documents, which makes additional manipulation
of any fielded bibliographic item (e.g. author list or bibcode or 
citation count) a very easy task using any vanilla XML parser.
Automated procedures that mimic the interactive session between a 
user and the ADS search engine can therefore be easily implemented
using a sequence of query-retrieve-parse steps.  For more information
on the tools and formats 
currently available please see the ADS on-line help pages.

Another feature recently added to the ADS which facilitates the
automatic parsing of bibliographic records is the availability of
RSS feeds for any ADS query.  At the bottom of the page containing
results from an abstract query users will find a link to the RSS
feed corresponding to the query results.  By using the link with
an RSS reader client or by parsing the content using a
parsing program one can immediately see if a new record satisfying
his or her query has been integrated in ADS.
Please note that since the
information transmitted in RSS feeds is quite limited, this approach
would not be suitable for compiling citation counts (which are not
included in the RSS streams).  However, this approach could be used
as the first step to discover any new bibliographic entries recently
added on a particular subject or by a particular author,
and then by retrieving the citations using follow-up queries.
This can be very useful to people maintaining bibliographic 
lists of papers about a mission or instrument.

\section{Conclusions}

We are often asked both by individual users and by librarians
what the coverage and completeness of citations are in the ADS.  
As we have discussed in this paper, 
there is no simple answer to this question since it very much 
depends on the particular area of interest one looks at.
Because the ADS has very good coverage
of the astronomy and core physics journals and has ongoing
collaborations with the main astronomy and physics publishers,
we can generally say that in the field
of astronomy and astrophysics the ADS provides a very good
account of the citations for the published literature.
The picture is not so clear when we start looking at papers
in other fields of physics, since they may be referenced by
articles in disciplines such as mathematics, 
chemistry, and computer science, which ADS does not cover well.
It is therefore natural for people to ask how well 
citations in ADS compare with the competition.
Our own analysis as well as independent   
studies comparing ADS and other sources 
\citep{2003lisa.conf..185S,gomez} have shown that 
citation coverage and depth have been improving in ADS over the 
past several years and that in most cases ADS gives the most complete 
citation results for researchers in astronomy.
We believe that the recent addition of citation data from e-prints
as well as our well-established relationship with
editors and users in the astronomical community will continue to
give us an edge over the other on-line abstracting services.

However, the use of citations in the ADS for bibliometric purposes 
is just one of its advantages.  
For the purpose of information discovery, 
the links between 
citing and cited articles are the primary purpose for maintaining
citations in the ADS.
The actual citation counts, while interesting, are a 
secondary by-product of the primary goal, which is to allow 
scientists to easily find and access those articles that will 
aid their research.  In this regard, the citation lists, 
along with article readership information,
provide us the necessary data for implementing the
powerful follow-up queries used by the 
second-order operators and by the {\em myADS} notification service.

\acknowledgements 

The ADS is funded by NASA Grant NNG06GG68G.


\begin{thebibliography}

\bibitem[Accomazzi et al.(1999)]{1999ASPC..172..291A} Accomazzi, A., 
Eichhorn, G., Kurtz, M.~J., Grant, C.~S., \& Murray, S.~S.\ 1999, ASP 
Conf.~Ser.~172: Astronomical Data Analysis Software and Systems VIII,
291 

\bibitem[Kurtz et al.(2002)]{2002SPIE.4847..238K} Kurtz, M.~J., Eichhorn, 
G., Accomazzi, A., Grant, C.~S., \& Murray, S.~S.\ 2002, 
Proceedings of the SPIE, 4847, 238 

\bibitem[Kurtz et al.(2005)]{2005IPM....41.1395K} Kurtz, M.~J., Eichhorn, 
G., Accomazzi, A., Grant, C., Demleitner, M., Henneken, E., \& Murray, 
S.~S.\ 2005, Information Processing and Management, 41, 1395 

\bibitem[Henneken et al.(2006)]{2006cs........8027H} Henneken, E., Kurtz, 
M.~J., Eichhorn, G., Accomazzi, A., Grant, C.~S., Thompson, D., Bohlen, E., 
\& Murray, S.~S.\ 2006, arXiv Computer Science e-prints, arXiv:cs/0608027 

\bibitem[Demleitner et al.(2004)]{2004ccdm.conf..521D} Demleitner, M., 
Kurtz, M., Accomazzi, A., Eichhorn, G., Grant, C.~S., \& Murray, S.~S.\ 
2004, Classification, Clustering, and Data Mining Applications.
Springer-Verlag, Berlin, 521 

\bibitem[G\'omez \& Merida-Martin (2007)]{gomez} G\'omez, M. \&
Merida-Martin, F.\ 2007, these proceedings

\bibitem[Grothkopf \& Stevens-Rayburn (2007)]{uta} Grothkopf, U. \&
Stevens-Rayburn, S.\ 2007, these proceedings

\bibitem[Kurtz et al.(2003)]{2003AAS...203.2005K} Kurtz, M.~J., Eichhorn, 
G., Accomazzi, A., Grant, C.~S., Henneken, E.~A., Thompson, D.~M., Bohlen, 
E.~H., \& Murray, S.~S.\ 2003, Bulletin of the American Astronomical 
Society, 35, 1241 

\bibitem[Stevens-Rayburn \& Bouton(2003)]{2003lisa.conf..185S} 
Stevens-Rayburn, S., \& Bouton, E.~N.\ 2003, Library and Information 
Services in Astronomy IV, 185 

\end{thebibliography}
\end{document}